\documentclass{elsart4-1}

\usepackage{amssymb}

\usepackage[english,francais]{babel}

\newtheorem{e-proposition}[theorem]{Proposition}

\newtheorem{e-definition}[theorem]{Definition\rm}

\renewcommand{\theequation}{\arabic{equation}}
\def\og{\leavevmode\raise.3ex\hbox{$\scriptscriptstyle\langle\!\langle$~}}
\def\fg{\leavevmode\raise.3ex\hbox{~$\!\scriptscriptstyle\,\rangle\!\rangle$}}

\begin{document}

\begin{frontmatter}


\selectlanguage{english}
\title{Nonlinear mean-field Fokker-Planck equations and their applications in physics, astrophysics and biology}
\vspace{-4cm} \selectlanguage{francais}
\title{ }


\selectlanguage{english}
\author[authorlabel1]{Chavanis Pierre-Henri}
\ead{chavanis@irsamc.ups-tlse.fr}

\address[authorlabel1]{Laboratoire de Physique Th\'eorique, Universit\'e Paul Sabatier, 118 route de Narbonne, 31062 Toulouse, France}

\begin{abstract}
We discuss a general class of nonlinear mean-field Fokker-Planck
equations [P.H. Chavanis, Phys. Rev. E, {\bf 68}, 036108 (2003)] and
show their applications in different domains of physics,
astrophysics and biology.  These equations are associated with
generalized entropic functionals and  non-Boltzmannian distributions
(Fermi-Dirac, Bose-Einstein, Tsallis,...). They furthermore involve
an arbitrary binary potential of interaction. We emphasize analogies
between different topics (two-dimensional turbulence,
self-gravitating systems, Debye-H\"uckel theory of electrolytes,
porous media, chemotaxis of bacterial populations, Bose-Einstein
condensation, BMF model, Cahn-Hilliard equations,...) which were
previously disconnected.  All these examples (and probably many
others) are particular cases of this general class of nonlinear
mean-field Fokker-Planck equations.

\end{abstract}
\end{frontmatter}

\selectlanguage{english}
\section{Introduction}
\label{intro}

Recently, several researchers have questioned the ``universality''
of the Boltzmann distribution in physics. This problem goes back to
Einstein himself who did not accept Boltzmann's principle $S=k\ln W$
on a general scope because he argued that the statistics of a system
$(W)$ should follow from its dynamics and cannot have a universal
expression \cite{pais}. In 1988, Tsallis introduced a more general
form of entropic functional in an attempt to describe complex systems
\cite{tsallis}. This was the starting point for several
generalizations of thermodynamics, statistical mechanics and kinetic
theories.  A lot of experimental and numerical studies have then
shown that complex media exhibit non-standard (non-Boltzmannian)
distributions and that Tsallis $q$-distributions can fit a wide
diversity of results when the Boltzmann distribution fails. However,
there also exists situations that are described neither by the
Tsallis nor by the Boltzmann distribution. The important point is to
explain {\it why} a given system exhibits non-Boltzmannian
distributions and find the reason for that. We cannot simply invoke
Tsallis entropy each time the system is ``complex''.

In a recent series of papers \cite{gfp}, we have argued that the
Tsallis distributions and the associated entropic functional are just
{\it particular} (yet important) types of non-Boltzmannian
distributions and entropies, and  that there is no fundamental
reason why they should play a prominent, or universal, role in
physics. Other distributions can emerge as well in complex systems
for different reasons.  On general grounds, we have suggested that
non-Boltzmannian distributions arise when ``hidden constraints'' are
in action and prevent the {\it a priori} accessible microstates to
be equiprobable. These constraints can be of different nature: (i)
There can be exclusion or inclusion constraints on the probabilities
of transition of the system from one state to the other. A
fundamental example is the Pauli exclusion principle in quantum
mechanics which prevents two fermions to occupy the same site in
phase space. This leads to the Fermi-Dirac statistics, instead of
the Boltzmann statistics, placing a bound $f({\bf r},{\bf v},t)\le
\eta_{0}$ on the distribution function. Similarly, one can account
for excluded volume constraints in classical systems of particles.  In
simplest models, this leads to the Fermi-Dirac statistics in physical
space placing a bound $\rho({\bf r},t)\le\sigma_{0}$ on the spatial
density. Alternatively, we can account for inclusion constraints (like
for bosons) by favoring the transition to a site that is already
occupied. This leads to the Bose-Einstein statistics. Nonlinear
Fokker-Planck equations taking into account these inclusion or
exclusion constraints have been discussed by Kaniadakis
\cite{kaniadakis}. (ii) We can imagine situations where some
microscopic constraints prevent the system from sampling the
energetically accessible phase space uniformly. These microscopic
constraints will modify the equilibrium distribution with respect to
the expected Boltzmann distribution. One example is the case of porous
media where the system has a complicated phase space structure
(fractal or multi-fractal) leading to anomalous diffusion.  The
Tsallis entropy \cite{tsallis} is adapted to this situation. The
dynamical evolution of the system can be described by a nonlinear
Fokker-Planck equation introduced by Plastino \& Plastino
\cite{pp}. (iii) In biology, the diffusion coefficient and the
mobility of the particles are often assumed to depend on the local
concentration so as to account for microscopic constraints that
cannot be easily modeled. This is the case for the Keller-Segel
model of chemotaxis \cite{keller} which can lead to non-standard
distributions at equilibrium. (iv) Finally, collisionless stellar
systems and inviscid 2D turbulent flows described by the
Vlasov-Poisson system and by the 2D Euler-Poisson system
spontaneously form meta-equilibrium states (galaxies and vortices)
that are usually described by non-Boltzmannian distributions
\cite{houches}. This is due, on the one hand, to the existence of an
infinite family of Casimir invariants for the fine-grained dynamics
which lead to non-standard distributions on the coarse-grained
scale. These Casimirs play the role of ``hidden constraints''
because they are not directly accessible from the coarse-grained
flow where the observations are made. Generalized Fokker-Planck
equations have been introduced by Robert \& Sommeria \cite{rsprl}
and Chavanis {\it et al.} \cite{csr}  to take into account these
constraints. On the other hand, collisionless mixing
 is incomplete in general and other forms of
distributions can emerge that are different from those predicted by
the statistical mechanics of violent relaxation.  These
quasi-stationary states (QSS) are nonlinearly dynamically stable
stationary solutions of the Vlasov or the 2D Euler equations
selected by the dynamics. In this context, the Tsallis distributions
are particular stationary solutions of the Vlasov equation known as
stellar polytropes in astrophysics. They can sometimes arise as a
result of incomplete violent relaxation, like in 2D turbulence, but
this is not generic \cite{brands}.  For collisionless stellar
systems and 2D vortices described by the Vlasov and the 2D  Euler
equations, non-standard distributions are explained by microscopic
constraints (the Casimirs) and incomplete violent relaxation
(non-ergodicity) \cite{next05}.

In this paper, we discuss a general class of nonlinear mean-field
Fokker-Planck equations associated with generalized entropic
functionals (leading to non-standard distributions) and an arbitrary
form of binary potential of interaction \cite{gfp}. These equations
are consistent with an effective generalized thermodynamical formalism
(E.T.F.).  The study of these equations is interesting in its own
right due to their rich mathematical properties but we provide here
explicit examples (2D turbulence, self-gravitating systems, porous
media, bacterial colonies, Bose-Einstein condensation...) where these
nonlinear mean-field Fokker-Planck equations can have physical
applications.

\section{Nonlinear mean-field Fokker-Planck equations}
\label{gmffp}

We consider a generalized class of nonlinear mean-field Kramers
equations of the form \cite{gfp}:
\begin{equation}
\label{gsk1}
{\partial f\over\partial t}+{\bf v}\cdot{\partial f\over\partial {\bf
r}}-\nabla\Phi\cdot{\partial f\over\partial {\bf
v}}={\partial\over\partial {\bf v}}\cdot\biggl\lbrace D\biggl\lbrack f
C''(f){\partial f\over\partial {\bf v}}+\beta f {\bf
v}\biggr\rbrack\biggr\rbrace,
\end{equation}
where $f=f({\bf r},{\bf v},t)$ is the distribution function, $C$ is a
convex function (i.e. $C''\ge 0$), $\beta=1/T$ is the inverse
temperature and $\Phi({\bf r},t)$ is a mean-field potential. The
friction parameter satisfies an Einstein relation $\xi=D\beta$. Since
the temperature $T$ is fixed, the above equation describes a canonical
situation. We assume that the potential is related to the density by a
general relation of the form
\begin{equation}
\label{gsk1b}
\Phi({\bf r},t)=\int u(|{\bf r}-{\bf r}'|)\rho({\bf r}',t)d{\bf r}',
\end{equation}
where $u$ is a binary potential of interaction and  $\rho=\int f d{\bf v}$ is the spatial density. We
introduce the functional
\begin{equation}
\label{gsk2}
F[f]={1\over 2} \int f v^{2}d{\bf r}d{\bf v}+{1\over 2}\int \rho
\Phi d{\bf r}+T \int C(f)d{\bf r}d{\bf v}.
\end{equation}
This functional can be interpreted as a free energy $F=E-TS$, where
$S$ is a ``generalized entropy'' and $E=K+W$ is the energy including a
kinetic term and a potential term. When $S[f]=-\int f\ln f d{\bf
r}d{\bf v}$ is the Boltzmann entropy, Eq. (\ref{gsk1}) reduces to the
ordinary Kramers equation. More generally, it is straightforward to
check that $F[f]$ plays the role of a Lyapunov functional satisfying
$\dot F\le 0$. This
is the equivalent, in the canonical ensemble, of the
H-theorem. The stationary solutions $f_{eq}({\bf r},{\bf v})$ are determined by the {\it integro-differential} equation
\begin{equation}
\label{gsk3}
C'(f_{eq})=-\beta \biggl \lbrack {v^{2}\over 2}+\Phi({\bf r}) \biggr \rbrack -\alpha,
\end{equation}
where $\Phi$ depends on $f_{eq}$ by Eq. (\ref{gsk1b}).   This
equation results from the {individual} cancelation of the advective
term (L.H.S.) and collision term (R.H.S.) in Eq. (\ref{gsk1})
\cite{gfp}. Since $C$ is convex, Eq. (\ref{gsk3}) can be inversed so that
$f_{eq}=f_{eq}(\epsilon)$ is a decreasing function of the individual
energy $\epsilon=v^{2}/2+\Phi({\bf r})$. The linearly dynamically
{\it stable} stationary solutions of Eq. (\ref{gsk1}) are  {\it
minima} of the free energy $F[f]$ at fixed mass \cite{gfp}.

In the strong friction limit $\xi\rightarrow +\infty$, or for large times $t\gg\xi^{-1}$, the distribution function  $f({\bf r},{\bf v},t)$ is
close to the {\it isotropic} distribution function determined by the
relation
\begin{equation}
\label{gsk4}
C'(f)=-\beta \biggl \lbrack {v^{2}\over 2}+\lambda({\bf r},t)\biggr \rbrack +O(\xi^{-1}).
\end{equation}
The pressure $p\equiv {1\over d}\int f v^{2} d{\bf v}=p\lbrack
\lambda({\bf r},t)\rbrack$ and the density $\rho\equiv \int f d{\bf
v}=\rho\lbrack \lambda({\bf r},t)\rbrack$ are related to each other
by a barotropic equation of state $p=p(\rho)$ which is entirely
specified by the function $C(f)$.  This equation of state is
obtained by eliminating $\lambda({\bf r},t)$ between the foregoing
relations. Taking the hydrodynamic moments of the generalized
Kramers equation (\ref{gsk1}) and closing the hierarchy by
considering the limit $\xi\rightarrow +\infty$, it is shown in
\cite{gfp} that the time evolution of the density $\rho({\bf r},t)$
is governed by the generalized Smoluchowski equation
\begin{equation}
\label{gsk5}
{\partial\rho\over\partial t}=\nabla\cdot\biggl \lbrack
{1\over\xi}(\nabla p+\rho\nabla\Phi)\biggr\rbrack.
\end{equation}
A formal derivation of the generalized Smoluchowski equation
(\ref{gsk5}) from the generalized Kramers equation (\ref{gsk1}) can
be realized by performing a Chapman-Enskog expansion in powers of
$\xi^{-1}$ \cite{lemou}.  The generalized Smoluchowski equation
monotonically decreases the Lyapunov functional
\begin{equation}
\label{gsk6}
F[\rho]= \int \rho\int^{\rho}{p(\rho')\over\rho^{'2}}d\rho'd{\bf r} + {1\over 2}\int\rho\Phi d{\bf r},
\end{equation}
which is the simplified form of free energy (\ref{gsk2}) obtained by
using the isotropic distribution function (\ref{gsk4}) to express
$F[f]$ as a functional of $\rho$ \cite{gfp,lemou}.  A stationary solution satisfies the
condition of hydrostatic equilibrium
\begin{equation}
\label{gsk7}
\nabla p+\rho\nabla\Phi={\bf 0}.
\end{equation}
Taking the hydrodynamic moments of the nonlinear mean-field Kramers
equation (\ref{gsk1}) and closing the hierarchy by a condition of
local thermodynamic equilibrium (in the canonical ensemble) one gets a
system of hydrodynamical equations
\begin{equation}
\label{gsk7a}
\frac{\partial\rho}{\partial t}+\nabla\cdot (\rho {\bf u})=0,
\end{equation}
\begin{equation}
\label{gsk7b} \frac{\partial {\bf u}}{\partial t}+({\bf u}\cdot
\nabla){\bf u}=-\frac{1}{\rho}\nabla p-\nabla\Phi-\xi{\bf u},
\end{equation}
that we called the damped barotropic Euler equations \cite{gfp}. In
the strong friction limit $\xi\rightarrow +\infty$, one can neglect
the inertial term in Eq. (\ref{gsk7b}) to obtain $\rho{\bf
u}=-\frac{1}{\xi}(\nabla p+\rho\nabla\Phi)+O(\xi^{-2})$.
Substituting this relation in the continuity equation (\ref{gsk7a}), we
recover the generalized Smoluchowski equation (\ref{gsk5}). The generalized
Smoluchowski equation can also be written
\begin{equation}
\label{hse2} {\partial \rho\over\partial t}=\nabla\cdot
\biggl\lbrace D\biggl\lbrack \rho C''(\rho)\nabla\rho+\beta
\rho\nabla\Phi\biggr \rbrack \biggr\rbrace.
\end{equation}
In the general formulation, $D$ can depend on ${\bf r}$ and $t$, so
on  $\rho({\bf r},t)$ \cite{gfp}.  The generalized Smoluchowski
equation (\ref{hse2}) monotonically decreases the generalized free energy
\begin{equation}
\label{hse4} F[\rho]=E-TS={1\over 2}\int \rho \Phi d{\bf r}+T\int
C(\rho)d{\bf r}.
\end{equation}
The stationary solutions of Eq. (\ref{hse2}) are solutions of the
integro-differential equation
\begin{equation}
\label{hse5} C'(\rho_{eq})=-\beta\Phi({\bf r})-\alpha,
\end{equation}
where $\Phi$ depends on $\rho_{eq}$ by Eq. (\ref{gsk1b}). Since $C$
is convex, Eq. (\ref{hse5}) can be inversed so that
$\rho_{eq}=\rho_{eq}(\Phi)$ is a decreasing function of the
potential (assuming $\beta>0$). The linearly dynamically {\it
stable} stationary solutions of Eq. (\ref{hse2}) are {\it minima} of
the free energy $F[\rho]$ at fixed mass \cite{gfp}.

These generalized Kramers and Smoluchowski equations can be obtained
in different ways: (i) from the linear thermodynamics of Onsager
\cite{frank,gfp} (ii) from a variational principle called Maximum
Entropy Production Principle (MEPP) adapted to the canonical
description \cite{gfp} (iii) from stochastic  Langevin equations
where the diffusion coefficient (noise) and the friction or mobility
coefficients explicitly depend on the density of particles
\cite{borland,gfp} (iv) from the Master equation, by allowing the
transition probabilities to depend on the population in the initial
and arrival states \cite{kaniadakis}. These generalized kinetic
equations can be justified as effective equations modeling
heuristically microscopic constraints that are not directly accessible
to the observer. It is clear that the mathematical study of these
equations is of considerable interest. A systematic study of the
generalized mean-field Smoluchowski equation (\ref{gsk5}) has been
undertaken in \cite{total,lang,iso,virial,hmf,bose} for different
equations of state $p(\rho)$, or generalized entropies $S[\rho]$, and
different potentials of interaction $u({\bf r}-{\bf r}')$ in different
dimensions of space $d$. Note that when the potential of interaction
is short-range, the mean-field potential is given by $\Phi\simeq
-\rho-R^{2}\Delta\rho$ and, when substituted in Eq. (\ref{hse2}), we get a
generalized form of the Cahn-Hilliard equation (see
\cite{gfp,lemou}). We now give explicit examples where these
nonlinear mean-field Fokker-Planck equations have physical applications.

\section{Two-dimensional turbulence}
\label{t}

Two-dimensional flows with high Reynolds numbers are described by the 2D Euler equations
\begin{equation}
{\partial \omega\over\partial t}+{\bf u}\cdot \nabla \omega=0,
\qquad {\bf u}=-\hat{\bf z}\times \nabla\psi, \qquad \omega=-\Delta\psi.
\label{t1}
\end{equation}
In geophysical fluid dynamics, one also considers the
Quasi-Geostrophic (QG) equations  where the vorticity $\omega$ is
replaced by the potential vorticity $q$ related to the
streamfunction $\psi$ by $q=-\Delta\psi+{1\over R^{2}}\psi$ where
$R$ is the Rossby radius. Starting from a generically unstable
initial condition ${\bf u}_{0}({\bf r})$, the 2D Euler equations are
known to develop a complicated mixing process which ultimately leads
to the emergence of a large-scale coherent structure, typically a
jet or a vortex. Jovian atmosphere shows a wide diversity of
structures (Jupiter's great red spot, white ovals, brown barges,
...). One question of fundamental interest is to predict the
structure and the stability of these ``equilibrium'' states
depending on the initial conditions. To that purpose, Robert \&
Sommeria \cite{mrs} have proposed a statistical mechanics of the 2D
Euler equation. The idea is to replace the deterministic evolution
of the flow $\omega({\bf r},t)$ by a probabilistic description where
$\rho({\bf r},\sigma,t)$ gives the density probability of finding
the vorticity level $\omega=\sigma$ in ${\bf r}$ at time $t$. The
observed (coarse-grained) vorticity field is then expressed as
$\overline{\omega}({\bf r},t)=\int \rho\sigma d\sigma$.

Consider first the case where the initial vorticity field consists
of an ensemble of patches with vorticity $\omega=\sigma_{0}$
surrounded by irrotational flow $\omega=0$. These patches will mix
in a complicated way but their vorticity and their total area
$\gamma_{0}$ will remain constant. In this two-levels approximation,
the conservation of the area is equivalent to the conservation of
the circulation $\Gamma=\int \overline{\omega} d{\bf
r}=\gamma_{0}\sigma_{0}$. The energy $E={1\over 2}\int
\overline{\omega}\psi d{\bf r}$ is also conserved. Using a result of
large deviations, it can be  shown \cite{mrs} that the optimal
probability $p({\bf r})$ of finding the level $\sigma_{0}$ in ${\bf
r}$ at equilibrium is obtained by maximizing the mixing entropy
\begin{equation}
S[p]=-\int \lbrace p\ln p+(1-p)\ln (1-p)\rbrace d{\bf r}, \label{t2}
\end{equation}
at fixed $\gamma_{0}=\int p d{\bf r}$ (or $\Gamma$) and $E$. This
expression of entropy can be obtained from a combinatorial analysis
taking into account the ``incompressibility'' of the vorticity
field. Using the method of Lagrange multipliers, the optimal
probability $p_{*}({\bf r})$, and consequently the coarse-grained
vorticity $\overline{\omega}=p_{*}\sigma_{0}$, is given by
\begin{equation}
\overline{\omega}({\bf r})={\sigma_{0}\over 1+\lambda e^{\beta\sigma_{0}\psi({\bf r})}}, \qquad (\lambda>0).
\label{t3}
\end{equation}
Note that $\overline{\omega}\le \sigma_{0}$, as the coarse-grained
vorticity locally averages over the levels $\omega=0$ and
$\omega=\sigma_{0}$. This constraint is similar to the Pauli
exclusion principle in quantum mechanics and this is why (\ref{t2})
and (\ref{t3}) look similar to the Fermi-Dirac entropy and
the Fermi-Dirac distribution. Non-standard (non-Boltzmannian)
distributions arise here because the patches (levels) cannot overlap
(or cannot be compressed).

Robert \& Sommeria \cite{rsprl} have proposed a dynamical model to
describe the relaxation of the system towards statistical
equilibrium. By using a Maximum Entropy Production Principle (MEPP),
they obtain in the two-levels case an equation of the form
\begin{equation}
{\partial \overline{\omega}\over\partial t}+{\bf u}\cdot \nabla \overline{\omega}=\nabla\cdot \lbrack D (\nabla\overline{\omega}+\beta(t)\overline{\omega}(\sigma_{0}-\overline{\omega})\nabla\psi)\rbrack ,\qquad \overline{\omega}=-\Delta\psi,
\label{t4}
\end{equation}
\begin{equation}
\beta(t)=-{\int D\nabla\overline{\omega}\cdot\nabla\psi d{\bf
r}\over \int
D\overline{\omega}(\sigma_{0}-\overline{\omega})(\nabla\psi)^{2}d{\bf
r}}, \qquad D({\bf r},t)\propto
\omega_{2}=\overline{\omega}(\sigma_{0}-\overline{\omega}).
\label{t5}
\end{equation}
This relaxation equation enters in the class of generalized mean-field
Fokker-Planck equations of Sec. \ref{gmffp}. It is associated with the
Fermi-Dirac entropy in physical space $S[\overline{\omega}]=-\int
\lbrace
(\overline{\omega}/\sigma_{0})\ln(\overline{\omega}/\sigma_{0})+(1-\overline{\omega}/\sigma_{0})\ln
(1-\overline{\omega}/\sigma_{0})\rbrace d{\bf r}$ and with a potential
$u(|{\bf r}-{\bf r}'|)=-{1\over 2\pi}\ln |{\bf r}-{\bf r}'|$.  Note,
however, that in the present case the inverse temperature $\beta(t)$
evolves with time according to Eq. (\ref{t5}) in order to conserve the
energy. Therefore, Eqs.  (\ref{t4})-(\ref{t5}) increase the entropy
$S$ at fixed $E$ and $\Gamma$. This corresponds to a microcanonical
description while the equations presented in Sec. \ref{gmffp}
correspond to a canonical description
\cite{gfp}.

In the general case, the optimal
vorticity distribution $\rho({\bf r},\sigma)$ maximizes the mixing
entropy
\begin{equation}
\label{t6} S\lbrack \rho\rbrack=-\int \rho\ln\rho \ d{\bf r}d\sigma,
\end{equation}
at fixed $E$, $\Gamma$ and $\Gamma_{n>1}=\int
\overline{\omega^{n}}d{\bf r}=\int \rho\sigma^{n}d\sigma d{\bf r}$.
The last constraints are equivalent to the  conservation of the
Casimirs or of the area $\gamma(\sigma)=\int \rho d{\bf r}$ of each
vorticity level. We must also account for the local normalization condition $\int \rho
d\sigma=1$. This yields the Gibbs
state
\begin{equation}
\label{t7}
\rho({\bf r},\sigma)={1\over Z({\bf r})}\chi(\sigma)  e^{-(\beta\psi+\alpha)\sigma}, \qquad Z=\int_{-\infty}^{+\infty}\chi(\sigma)e^{-(\beta\psi+\alpha)\sigma}d\sigma,
\end{equation}
where $\chi(\sigma)={\rm exp}(-\sum_{n>1}\alpha_{n}\sigma^{n})$
encapsulates the Lagrange multipliers $\alpha_{n>1}$ accounting for
the conservation of the fragile constraints $\Gamma_{n>1}$ while
$\beta$ and $\alpha$ account for the conservation of the robust
constraints $E$ and $\Gamma$ ($Z$ is the normalization constant
determined by $\int \rho d\sigma=1$)\footnote{$E$ and $\Gamma$ are
called {\it robust constraints} because they can be expressed in
terms of the coarse-grained field. By contrast, the $\Gamma_{n>1}$
are called {\it fragile constraints} because they are not conserved
by the coarse-grained vorticity as $\overline{\omega^{n}}\neq
\overline{\omega}^{n}$ and they must be expressed in terms of the
fine-grained vorticity distribution $\rho({\bf r},\sigma)$. This
implies that $E$ and $\Gamma$ can be computed at any time from the
coarse-grained flow while the $\Gamma_{n>1}$ are only accessible
from the initial conditions or from the fine-grained flow. Since
they are not accessible from the macroscopic dynamics, they behave
as ``hidden constraints''.}. The coarse-grained vorticity
$\overline{\omega}=\int \rho \sigma d\sigma$ is now given by
\begin{equation}
\label{t8}
\overline{\omega}={\int \chi(\sigma)\sigma e^{\sigma(\beta\psi+\alpha)}d\sigma\over \int \chi(\sigma) e^{\sigma(\beta\psi+\alpha)}d\sigma}=F(\beta\psi+\alpha)=f(\psi),
\end{equation}
where $F(\Phi)=-(\ln \hat{\chi})'(\Phi)$ and $\hat{\chi}(\Phi)=\int_{-\infty}^{+\infty}
\chi(\sigma)e^{-\sigma\Phi}d\sigma$. In
this approach, $\chi(\sigma)$ is determined {\it a posteriori} by the
initial conditions through the conservation of the Casimirs. Indeed,
the $\alpha_{n}$ are Lagrange multipliers which must be related to
the constraints $\Gamma_{n}$. The general relaxation equations have the form
\begin{equation}
{\partial \rho\over\partial t}+{\bf u}\cdot \nabla \rho=\nabla\cdot
\lbrack D (\nabla\rho+\beta(t)\rho
(\sigma-\overline{\omega})\nabla\psi)\rbrack, \qquad D\propto
\omega_{2}=\int \rho (\sigma-\overline{\omega})^{2}d{\bf r},
\label{t9}
\end{equation}
where the time evolution of $\beta(t)$ is determined by the energy
constraint \cite{rsprl}.  These equations for $\rho({\bf r},\sigma,t)$ conserve
all the Casimirs (in addition to the circulation and the energy) and
increase the mixing entropy (\ref{t6}) until the maximum entropy
state (\ref{t7}) is reached. Note also that the diffusion
coefficient is not constant but depends on the local fluctuations of
vorticity. This can block the relaxation in a sub-region of the flow
(maximum entropy bubble) and account for incomplete relaxation
\cite{rsprl,csr}.

In the case of flows that are forced at small-scales, Ellis et al.
\cite{ellis} and Chavanis \cite{physD} have proposed to treat the
constraints associated with the fragile moments $\Gamma_{n>1}$
canonically and fix the  Lagrange multipliers $\alpha_{n>1}$, or the
{\it prior} distribution $\chi(\sigma)$, instead of $\Gamma_{n>1}$
(by contrast, the robust constraints $\Gamma$ and $E$ are still
treated microcanonically). If we view the vorticity levels as
species of particles, this amounts to fixing the chemical potentials
instead of the total number of particles in each species. It is
argued that the prior $\chi(\sigma)$ is imposed by the small-scale
forcing so it must be regarded as given {\it a priori} (it is
external to the system under consideration). In this point of view,
one must work with the relative entropy
\begin{equation}
\label{t10} S_{\chi}\lbrack \rho\rbrack=-\int \rho\ln\biggl\lbrack
{\rho\over \chi(\sigma)}\biggr\rbrack \ d{\bf r}d\sigma,
\end{equation}
where $\chi(\sigma)$ is assumed given. The relative entropy
(\ref{t10}) can be viewed as the Legendre transform
$S_{\chi}=S-\sum_{n>1}\alpha_{n}\Gamma_{n}$ of the mixing entropy
(\ref{t6}) with respect to the fragile moments $\Gamma_{n>1}=\int
\overline{\omega^{n}}d{\bf r}$ \cite{physD}. The equilibrium state
is obtained by maximizing (\ref{t10}) while conserving only the
robust constraints $E$ and $\Gamma$. This again yields the Gibbs
state (\ref{t7}) except that now $\chi(\sigma)$ is fixed {\it a
priori} by the small-scale forcing. The optimal coarse-grained
vorticity is given by (\ref{t8}) where $F(\Phi)=-(\ln
\hat{\chi})'(\Phi)$ is now completely specified by the prior
$\chi(\sigma)$. Note that $\overline{\omega}$ satisfies a result of
large deviations \cite{ellis}. It maximizes a generalized entropy
\cite{physD}:
\begin{equation}
\label{t11} S\lbrack \overline{\omega}\rbrack =-\int
C(\overline{\omega})  d{\bf r},
\end{equation}
at fixed circulation and energy, where $C$ is a convex function
directly determined by the prior according to the relation
\cite{super}:
\begin{equation}
\label{t12} C(\overline{\omega})=-\int^{\overline{\omega}}F^{-1}(x)dx=-\int^{\overline{\omega}}\lbrack (\ln {\hat \chi})'\rbrack^{-1}(-x)dx.
\end{equation}
In this context, Chavanis \cite{gfp,physD} has proposed a
thermodynamical small-scale parameterization of 2D forced geophysical
flows where the coarse-grained vorticity evolves according to
\begin{equation}
{\partial \overline{\omega}\over\partial t}+{\bf u}\cdot \nabla
\overline{\omega}=\nabla\cdot \biggl \lbrace D
\biggl\lbrack \nabla\overline{\omega}+{\beta(t)\over
C''(\overline{\omega})}\nabla\psi\biggr\rbrack\biggr\rbrace ,\qquad
\overline{\omega}=-\Delta\psi,
\label{t13}
\end{equation}
\begin{equation}
\beta(t)=-{\int D\nabla\overline{\omega}\cdot\nabla\psi d{\bf
r}\over \int D{(\nabla\psi)^{2}\over C''(\overline{\omega})}d{\bf
r}}, \qquad D\propto \omega_{2}={1\over C''(\overline{\omega})}.
\label{t14}
\end{equation}
This relaxation equation for $\overline{\omega}({\bf r},t)$ conserves
only the robust constraints (circulation and energy) and increases the
generalized entropy (\ref{t11}) fixed by the prior vorticity
distribution $\chi(\sigma)$. It differs from the relaxation equations
(\ref{t9}) of Robert \& Sommeria in the sense that the specification
of the prior $\chi(\sigma)$ (determined by the small-scale forcing)
replaces the specification of the Casimirs (determined by the initial
conditions). However, in both models, the robust constraints $E$ and
$\Gamma$ are treated microcanonically (i.e. they are rigorously
conserved). We thus have to solve {\it one} differential equation
(\ref{t13}) instead of $N$ coupled equations (\ref{t9}) for each
discretized level $\sigma_k$. The relaxation equation (\ref{t13})
enters in the class of generalized mean-field Fokker-Planck equations
presented in Sec. \ref{gmffp}. It is however associated with a
microcanonical description since $\beta(t)$ varies in time so as to
conserve the energy $E$ \cite{gfp}.

Because of incomplete relaxation, the coarse-grained vorticity can
converge towards a metaequilibrium state which is different from that
predicted by the statistical theory \cite{next05}. This coarse-grained
vorticity field is a stable stationary solution of the 2D Euler
equation which is incompletely mixed. The 2D Euler equation admits an
infinite number of stationary solutions specified by an arbitrary
$\overline{\omega}-\psi$ relationship. A solution of the form
$\overline{\omega}=f(\psi)$, where $f$ is monotonic, which maximizes
an $H$-function $H[\overline{\omega}]=-\int C(\overline{\omega})d{\bf
r}$ (where $C$ is convex) at fixed energy $E$ and circulation $\Gamma$
is nonlinearly dynamically stable with respect to the 2D Euler
equation \cite{ellis,physD}. For example, Tsallis functional
$H_{q}=-{1\over q-1}\int
(\overline{\omega}^{q}-\overline{\omega})d{\bf r}$ is a particular
$H$-function. Its maximization at fixed $E$ and $\Gamma$ determines a
particular class of stationary solutions of the 2D Euler equations
(polytropic vortices) corresponding to $\overline{\omega}=\lbrack
\lambda-\beta(q-1)\psi/q\rbrack^{1/(q-1)}$. Minus the enstrophy
$\Gamma_{2}=\int \overline{\omega}^{2}d{\bf r}$ can also be viewed as
an $H$-function leading to a linear $\overline{\omega}-\psi$
relationship. This is a particular case of the Tsallis functional
corresponding to $q=2$.  Note that the relaxation equations
(\ref{t13}) and (\ref{t14}) can serve as a {\it numerical algorithm} to
construct nonlinearly dynamically stable stationary solutions of the
2D Euler equation since the steady state of these equations maximizes,
by construction, an $H$-function specified by $C(\overline{\omega})$
at fixed circulation and energy.

\section{Collisionless stellar systems}
\label{ss}

For most stellar systems, the encounters between stars are
negligible \cite{bt} so that the galactic dynamics is described by
the Vlasov-Poisson system
\begin{equation}
{\partial f\over\partial t}+{\bf v}\cdot{\partial f\over\partial
{\bf r}}-\nabla\Phi\cdot{\partial f\over\partial {\bf v}}=0, \qquad
\Delta\Phi=4\pi G\int f d{\bf v}. \label{c1}
\end{equation}
The Vlasov-Poisson and the 2D Euler-Poisson systems present deep
similarities as shown in \cite{houches}. In particular, the Vlasov-Poisson
system develops a complicated mixing process in phase space
associated with the damped oscillations of a proto-galaxy initially
out-of-equilibrium and in search of a steady (virialized) state. Due
to ``phase mixing'', a collisionless stellar system can reach an
``equilibrium'' state (metaequilibrium) on a very short timescale.
This collisionless relaxation has been called violent relaxation.
Lynden-Bell \cite{lb} proposed a statistical mechanics of this
process which is similar to that exposed previously in 2D
turbulence. However, this process takes place in a six-dimensional
phase space instead of the plane \cite{csr}.

In the two-levels approximation where  $f=\eta_{0}$ or $f=0$, the optimal probability $p({\bf r},{\bf v})$ of finding the level $\eta_{0}$ in ${\bf r},{\bf v}$ at equilibrium is obtained by maximizing the mixing entropy
\begin{equation}
S[p]=-\int \lbrace p\ln p+(1-p)\ln (1-p)\rbrace d{\bf r} d{\bf v},
\label{c2}
\end{equation}
at fixed $\gamma_{0}=\int p d{\bf r}d{\bf v}$ (or total mass $M$)
and energy $E={1\over 2}\int \overline{f} v^{2} d{\bf r}d{\bf
v}+{1\over 2}\int \rho \Phi d{\bf r}$. Using the method of Lagrange
multipliers, the optimal probability $p_{*}({\bf r},{\bf v})$, and
consequently the coarse-grained distribution function
$\overline{f}=p_{*}\eta_{0}$, is given by
\begin{equation}
\overline{f}({\bf r},{\bf v})={\eta_{0}\over 1+\lambda e^{\beta\eta_{0}({v^{2}\over 2}+\Phi)}}.
\label{c3}
\end{equation}
Apart from a reinterpretation of the constants, this is similar to
the Fermi-Dirac distribution \cite{lb}. Using the analogy with 2D
turbulence, Chavanis et al. \cite{csr} have proposed a dynamical
model to describe the relaxation of the system towards equilibrium.
By using the MEPP, they obtain in the two-levels case an equation of
the form
\begin{equation}
{\partial \overline{f}\over\partial t}+{\bf v}\cdot{\partial
\overline{f}\over\partial {\bf r}}-\nabla\Phi\cdot{\partial
\overline{f}\over\partial {\bf v}}={\partial\over\partial {\bf
v}}\cdot\biggl\lbrace D\biggl\lbrack {\partial
\overline{f}\over\partial {\bf v}}+\beta(t) \overline{f}
(\eta_{0}-\overline{f}) {\bf v}\biggr\rbrack\biggr\rbrace, \qquad
\Delta\Phi=4\pi G\int \overline{f} d{\bf v}, \label{c4}
\end{equation}
\begin{equation}
\beta(t)=-{\int D {\partial \overline{f}\over\partial {\bf v}}\cdot
{\bf v} d{\bf r} d{\bf v}\over \int
D\overline{f}(\eta_{0}-\overline{f})v^{2}d{\bf r} d{\bf v}}.
\label{c5}
\end{equation}
This relaxation equation enters in the class of generalized
mean-field Fokker-Planck equations of Sec. \ref{gmffp} associated
with the Fermi-Dirac entropy in phase space $S[\overline{f}]=-\int
\lbrace
(\overline{f}/\eta_{0})\ln(\overline{f}/\eta_{0})+(1-\overline{f}/\eta_{0})\ln
(1-\overline{f}/\eta_{0})\rbrace d{\bf r}d{\bf v}$ and with a
Newtonian potential $u=-G/|{\bf r}-{\bf r}'|$. Note, however, that
in the present case the inverse temperature evolves with time
according to Eq. (\ref{c5}) so as to conserve the energy. Therefore,
Eq. (\ref{c4}) increases the entropy $S$ at fixed $E$ and $M$. This
describes a microcanonical situation \cite{gfp}.

In the general case, we get a system of relaxation equations for
$\rho({\bf r},{\bf v},\eta,t)$ of the form \cite{csr}:
\begin{equation}
{\partial \rho\over\partial t}+{\bf v}\cdot{\partial \rho\over\partial {\bf
r}}-\nabla\Phi\cdot{\partial \rho\over\partial {\bf
v}}={\partial\over\partial {\bf v}}\cdot\biggl\lbrace D\biggl\lbrack {\partial \rho\over\partial {\bf v}}+\beta(t) \rho (\eta-\overline{f}) {\bf
v}\biggr\rbrack\biggr\rbrace.
\label{c6}
\end{equation}
They converge at equilibrium towards the Gibbs state
\begin{equation}
\label{c7}
\rho({\bf r},{\bf v},\eta)={1\over Z({\bf r},{\bf v})}\chi(\eta)  e^{-(\beta\epsilon+\alpha)\eta},
\end{equation}
which maximizes the mixing entropy
\begin{equation}
\label{c8} S\lbrack \rho\rbrack=-\int \rho\ln\rho \ d{\bf r}d{\bf
v}d\eta,
\end{equation}
at fixed $E$, $M$ and $\gamma(\eta)=\int \rho d{\bf r}d{\bf v}$. At
statistical equilibrium, the coarse-grained distribution function
$\overline{f}=\int \rho \eta d\eta$ is given by
\begin{equation}
\label{c9}
\overline{f}={\int \chi(\eta)\eta e^{\eta(\beta\epsilon+\alpha)}d\eta\over \int \chi(\eta) e^{\eta(\beta\epsilon+\alpha)}d\eta}=F(\beta\epsilon+\alpha)=\overline{f}(\epsilon).
\end{equation}
We note that the statistical theory of violent relaxation leads to
non-standard (non-Boltzmannian) coarse-grained distribution
functions. This is due to the existence of an infinite family of
conserved quantities, the Casimirs, which play the role of ``hidden
constraints'' in our general interpretation of non-standard
distributions. The coarse-grained distribution functions arising in
theories of violent relaxation can be viewed as sorts of {\it
superstatistics} \cite{super} since they are expressed as
superpositions of Boltzmann distributions weighted by a
non-universal factor depending on the initial conditions.

Like in 2D turbulence, the violent relaxation of collisionless
stellar systems is usually incomplete \cite{lb}.  The Vlasov
equation admits an infinite number of stationary solutions specified
by the Jeans theorem.  A solution of the form
$\overline{f}={f}(\epsilon)$ with $f'(\epsilon)<0$ which maximizes
an $H$-function $H[\overline{f}]=-\int C(\overline{f})d{\bf r}d{\bf
v}$ (where $C$ is convex) at fixed energy $E$ and mass $M$ is
nonlinearly dynamically stable with respect to the Vlasov-Poisson
system \cite{nl}. This determines a subclass of spherical stellar systems. For
example, the Tsallis functional $H_{q}=-{1\over q-1}\int
(\overline{f}^{q}-\overline{f})d{\bf r}d{\bf v}$ is a particular
$H$-function. Its maximization at fixed $E$ and $M$ determines a
particular class of stationary solutions of the Vlasov-Poisson
system with DF $\overline{f}=\lbrack
\lambda-\beta(q-1)\epsilon/q\rbrack^{1/(q-1)}$ called stellar
polytropes in astrophysics (the standard polytropic index is related
to the $q$ parameter by $n=3/2+1/(q-1)$). Like in 2D turbulence, we
can use generalized Fokker-Planck equations as {\it numerical algorithms}
to construct stable stationary solutions of the Vlasov equation (see
\cite{gfp}).

\section{Self-gravitating Brownian particles and related systems}
\label{b}

In a series of papers, Chavanis \& Sire \cite{total} have introduced
and studied a model of self-gravitating Brownian particles.  In this
model, the particles interact via self-gravity but, in addition,
they experience a friction force (against an inert medium) and a
stochastic force. This Brownian model is described by the canonical
ensemble in contrast to the usual Hamiltonian model of stellar
systems which is described by the microcanonical ensemble. In the
mean-field limit, the evolution of the distribution function of the Brownian
gas is governed by the Kramers-Poisson system. In the strong
friction limit, we get the Smoluchowski-Poisson (SP) system
\begin{equation}
\label{gsk5we} {\partial\rho\over\partial t}=\nabla\cdot\biggl \lbrack
{1\over\xi}(T\nabla \rho+\rho\nabla\Phi)\biggr\rbrack,
\end{equation}
\begin{equation}
\label{hse2rf}\Delta\Phi=S_d G\rho.
\end{equation}
Comparing with Eq. (\ref{gsk5}), this corresponds to an isothermal equation
of state $p=\rho T$ associated with the Boltzmann entropy $S=-\int
\rho\ln\rho d{\bf r}$. In $d\ge 2$, there exists a critical
temperature $T_{c}$ for box-confined systems. For $T>T_{c}$, the
system reaches a stable equilibrium state corresponding to the
Boltzmann statistics
\begin{equation}
\label{hse2li}\rho_{eq}=Ae^{-\beta\Phi}.
\end{equation}
For $T<T_{c}$, there is no equilibrium state anymore. The density
blows up and finally forms a Dirac peak (condensate)
\cite{total}. Chavanis \& Sire \cite{lang} have also studied more
general models where the diffusion of particles is anomalous so that
the evolution of the particles is described by the nonlinear
Smoluchowski-Poisson (NSP) system
\begin{equation}
\label{gsk5frtrg}{\partial\rho\over\partial t}=\nabla \cdot \biggl\lbrack
{1\over\xi}(K\nabla \rho^{\gamma}+\rho\nabla\Phi)\biggr\rbrack,
\end{equation}
\begin{equation}
\label{hse2vf}\Delta\Phi=S_d G\rho.
\end{equation}
Comparing with Eq. (\ref{gsk5}), this corresponds to a polytropic equation of
state $p=K\rho^{\gamma}$ associated with the Tsallis entropy
$S=-{1\over\gamma-1}\int (\rho^{\gamma}-\rho)d{\bf r}$. The steady
states (polytropes) reproduce the statistics introduced by Tsallis
\begin{equation}
\label{hse2yt}\rho_{eq}=\left\lbrack \lambda-{\beta(q-1)\over q}\Phi\right \rbrack^{1\over q-1}.
\end{equation}
The nonlinear Smoluchowski equation (\ref{gsk5frtrg}) also arises in
the physics of porous media where anomalous diffusion is due to the
complicated (fractal, multi-fractal) phase-space structure of the
medium and $\Phi$ is an external potential \cite{pp}. More generally,
we can consider the evolution of particles described by the
generalized Smoluchowski-Poisson (GSP) system
\begin{equation}
\label{gsk5vfvd}{\partial\rho\over\partial t}=\nabla \cdot \biggl\lbrack
{1\over\xi}(\nabla p+\rho\nabla\Phi)\biggr\rbrack,
\end{equation}
\begin{equation}
\label{hse2reg}\Delta\Phi=S_d G\rho.
\end{equation}
The case of self-gravitating Brownian fermions where $p=p(\rho)$ is
the Fermi-Dirac equation of state has been studied in \cite{iso}.
Some general properties of the GSP system, including the Virial
theorem and a stability analysis of the solutions have been obtained
in \cite{virial}.

We can also consider other forms of potential of interaction than
the gravitational one. For example, the Smoluchowski-Poisson system
with the repulsive Coulombian potential describes electrolytes in
the theory of Debye-H\"uckel \cite{dh}. On the other hand, when the
particles evolve on a ring and the  potential of interaction is
truncated to one Fourier mode, we get the Brownian Mean Field (BMF)
model \cite{hmf}:
\begin{equation}
\label{hse2ab} {\partial\rho\over\partial
t}=T{\partial^{2}\rho\over\partial\theta^{2}}+{k\over
2\pi}{\partial\over\partial\theta}\biggl\lbrace\rho\int_{0}^{2\pi}\sin(\theta-\theta')\rho(\theta',t)d\theta'\biggr\rbrace,
\end{equation}
which is the canonical version of the Hamiltonian Mean Field (HMF)
model.

Finally, a dynamical model of the Bose-Einstein condensation has
been studied recently by Sopik et al. \cite{bose}. The dynamical
evolution of a gas of non-interacting bosons in contact with a
thermal bath is governed by the bosonic Kramers equation
\begin{equation}
\label{gsk5df} {\partial\rho\over\partial t}={1\over\xi}\nabla_{\bf
k}\cdot\biggl\lbrack  T\nabla_{\bf k} \rho+\rho (1+\rho) {\bf
k}\biggr\rbrack,
\end{equation}
which takes into account an inclusion principle \cite{kaniadakis}.
This equation enters in the generalized class of Fokker-Planck
equations of Sec. \ref{gmffp}. It is associated with the Bose-Einstein
entropy $S=-\int \lbrace \rho\ln\rho-(1+\rho)\ln (1+\rho)\rbrace d{\bf
k}$. We have found that the Bose-Einstein condensation (in phase
space) shares deep similarities with the collapse of the
self-gravitating Brownian gas (in position space). In particular, for
$d\ge 3$, there exists a critical temperature $T_c$. The density blows
up for $T<T_c$ leading ultimately to a Bose condensate in ${\bf
k}$-space \cite{bose}.

\section{Chemotactic aggregation of bacterial populations}
\label{chem}

 The name
chemotaxis refers to the motion of organisms (amoeba) induced by
chemical signals (acrasin). In some cases, the biological organisms
secrete a substance that has an attractive effect on the organisms
themselves. Therefore, in addition to their diffusive motion, they
move systematically along the gradient of concentration of the
chemical they secrete (chemotactic flux). When attraction prevails
over diffusion, the chemotaxis can trigger a self-accelerating
process until a point at which aggregation takes place. This is the
case for the slime mold {\it Dictyostelium Discoideum} and for the
bacteria {\it Escherichia coli}.

A model of slime mold aggregation has been introduced by Keller \&
Segel \cite{keller} in the form of two PDE's:
\begin{equation}
\label{ca1} {\partial\rho\over\partial t}=\nabla\cdot (D_{2}\nabla\rho)-\nabla\cdot (D_{1}\nabla c),
\end{equation}
\begin{equation}
\label{ca2}{\partial c\over\partial t}=-k(c)c+f(c)\rho+D_{c}\Delta c.
\end{equation}
In these equations $\rho({\bf r},t)$ is the concentration of amoebae and
$c({\bf r},t)$ is the concentration of acrasin. Acrasin is produced
by the amoebae at a rate $f(c)$. It can also be degraded at a rate
$k(c)$. Acrasin diffuse according to Fick's law with a diffusion
coefficient $D_c$. Amoebae concentration changes as a result of an
oriented chemotactic motion in the direction of a positive gradient of
acrasin and a random motion analogous to diffusion. In Eq. (\ref{ca1}),
$D_2(\rho,c)$ is the diffusion coefficient of the amoebae and
$D_1(\rho,c)$ is a measure of the strength of the influence of the
acrasin gradient on the flow of amoebae. This chemotactic drift is the
fundamental process in the problem.

A first simplification of the Keller-Segel model is provided by
the system of equations
\begin{equation}
\label{ca3} {\partial\rho\over\partial t}=D\Delta\rho-\chi\nabla\cdot (\rho\nabla c),
\end{equation}
\begin{equation}
\label{ca4}{\partial c\over\partial t}=D'\Delta c+a \rho-b c,
\end{equation}
where the parameters are positive constants. An additional
simplification, introduced by J\"ager \& Luckhaus \cite{jager},
consists in ignoring the temporal derivative in Eq. (\ref{ca4}). This is
valid in the case where the diffusion coefficient $D'$ is
large. Taking also $b=0$, we obtain
\begin{equation}
\label{ca5} {\partial\rho\over\partial t}=D\Delta\rho-\chi\nabla \cdot (\rho\nabla c),
\end{equation}
\begin{equation}
\label{ca6}\Delta c=-\lambda\rho,
\end{equation}
where $\lambda=a/D'$. These equations are isomorphic to the
Smoluchowski-Poisson system (\ref{gsk5we})-(\ref{hse2rf}) which
describes self-gravitating Brownian particles in a high friction
limit. The analogy between self-gravitating systems and the chemotaxis
of bacterial colonies is developed in \cite{iso}.

The Keller-Segel
model ignores clumping and sticking effects. However, at the late
stages of the blow-up, when the density of amoebae has reached high
values, finite size effects and stickiness must clearly be taken into
account. As a first step, we have proposed in \cite{iso} to replace the
classical equation (\ref{ca5}) by an equation of the form
\begin{equation}
\label{gcm1} {\partial\rho\over\partial t}=D\Delta\rho-\chi\nabla \cdot (\rho(1-\rho/\sigma_{0})\nabla c),
\end{equation}
which enforces a limitation $\rho\le \sigma_{0}$ on the maximum
concentration of bacteria in physical space. This equation decreases
the Lyapunov functional
\begin{equation}
\label{gcm2} {F}[\rho]=-{1\over 2}\int \rho c  \ d{\bf r}+{D\sigma_{0}\over \chi}\int \left\lbrack {\rho\over\sigma_{0}}\ln{\rho\over\sigma_{0}}+\left (1-{\rho\over\sigma_{0}}\right )\ln\left (1-{\rho\over\sigma_{0}}\right )\right \rbrack d{\bf r}.
\end{equation}
This functional can be interpreted as a free energy $F=E-T_{eff}S$
associated with a Fermi-Dirac entropy in physical space
\cite{gfp,iso} where $T_{eff}=D/\chi$ is an effective temperature.
This form of entropy can be obtained by introducing a lattice model
preventing two particles to be on the same site. The lattice creates
an exclusion principe in physical space similar to the Pauli
exclusion principle in phase space for fermions. Then, $S[\rho]$ can
be obtained by a standard combinatorial analysis respecting this
exclusion principle. The equilibrium states of Eq. (\ref{gcm1}) are
given by a Fermi-type distribution in physical space
\begin{equation}
\label{gcm3} \rho={\sigma_{0}\over 1+\lambda e^{-{\chi\over D}c}},\qquad (\lambda>0),
\end{equation}
which minimizes the effective free energy (\ref{gcm2}) at fixed
mass. On the other hand, the relation between the concentration of
amoebae and acrasin may be more complex that simply given by the
Poisson equation (\ref{ca6}). For example, taking $b\neq 0$ in the
original Keller-Segel model, we obtain a relation of the form
\begin{equation}
\label{gcm4}\Delta c-k_{S}^{2}c=-\lambda\rho,
\end{equation}
where $k_{S}^{2}=b/D'$.  The second term is similar to the Debye
shielding in plasma physics or to the Rossby shielding in
Quasi-Geostrophic (QG) flows. We note that Eq. (\ref{gcm1}) enters
in the class of nonlinear mean-field  Fokker-Planck equations of
Sec. \ref{gmffp} associated with a Fermi-Dirac entropy in physical
space. We also note the resemblance with the relaxation equations
(\ref{t4}) of 2D turbulence except that here $\beta\equiv \chi/D$ is
fixed (which is similar to a ``canonical'' situation) while in 2D
turbulence $E$ is fixed (microcanonical situation).

More generally, in the primitive Keller-Segel model
(\ref{ca1})-(\ref{ca2}) the coefficients $D_{1}$ and $D_{2}$ can both
depend on the concentration of particles. This takes into account
``hidden constraints'' that act on the microscopic dynamics.  Equation
(\ref{gcm1}) is an example of nonlinear mean-field Fokker-Planck
equations where the mobility depends on the concentration while the
diffusion coefficient is constant. Alternatively, if the diffusion
coefficient depends on the concentration while the mobility is
constant, we can write Eq. (\ref{ca1}) in the form
\begin{equation}
\label{gsk5vfre}{\partial\rho\over\partial t}=\nabla \cdot \left\lbrack
\chi (\nabla p-\rho\nabla c)\right\rbrack,
\end{equation}
where $p=p(\rho)$ is an effective pressure. Therefore, the
Keller-Segel model is an example of nonlinear mean-field
Fokker-Planck equations presented in Sec. \ref{gmffp}. We can thus develop an
Effective Thermodynamical Formalism (ETF) to investigate the
chemotactic problem. We can also introduce
  an inertial model of chemotaxis \cite{iso,virial} based on the
hydrodynamic equations
\begin{equation}
\label{gsk7c}
\frac{\partial\rho}{\partial t}+\nabla\cdot (\rho {\bf u})=0,
\end{equation}
\begin{equation}
\label{gsk7d} \frac{\partial {\bf u}}{\partial t}+({\bf u}\cdot
\nabla){\bf u}=-\frac{1}{\rho}\nabla p+\nabla c-\xi{\bf u},
\end{equation}
\begin{equation}
\label{ca2w}{\partial c\over\partial t}=-k(c)c+f(c)\rho+D_{c}\Delta
c.
\end{equation}
These hydrodynamical equations can be derived from generalized
Fokker-Planck equations of the form (\ref{gsk1}) \cite{virial}. They
are similar to the damped Euler equations (\ref{gsk7a})-(\ref{gsk7b})
for self-gravitating Brownian particles \cite{gfp} except that the
Poisson equation is replaced by a more general field equation taking
into account the specificities of the chemotactic model. For $\xi=0$,
we recover the hyperbolic model introduced by Gamba et
al. \cite{gamba} and for $\xi\rightarrow +\infty$, we can neglect the
inertial term in Eq. (\ref{gsk7d}), leading to $\rho{\bf
u}=-\frac{1}{\xi}(\nabla p+\rho\nabla\Phi)+O(\xi^{-2})$. When
substituted in Eq. (\ref{gsk7c}), we recover the parabolic
Keller-Segel model (\ref{gsk5vfre}). It has been shown that parabolic
models of chemotaxis lead to point-wise blow-up while hyperbolic
models show the formation of a network that is interpreted as the
beginning of a vasculature.  This phenomenon is responsible of
angiogenesis, a major factor for the growth of tumors.

\section{Conclusion}
\label{}

In this paper, we have given several physical examples where the
evolution of the system is governed by nonlinear mean-field
Fokker-Planck equations introduced in \cite{gfp}. These equations
are consistent with an Effective Thermodynamical Formalism (ETF) and
lead to non-standard distributions at equilibrium. We stress that:
(i) Tsallis distributions and Tsallis generalized entropies just
represent a particular case of this general formalism. (ii) The
physical reason why non-standard distributions arise in a problem is
different from case to case and must be sought for each particular
system. In general, this reflects the existence of ``hidden
constraints'' \cite{gfp} that can be of various forms as discussed
in the Introduction. If the system does not ``mix well'' (for some
reason) there is no universal form of entropy to account for
non-ergodicity. (iii) In many cases, the resemblance with a
``generalized thermodynamical formalism'' is essentially effective
or formal. In some cases, there is no relation with thermodynamics
at all. Yet, it may be of interest to develop a {\it thermodynamical
analogy} and use the same vocabulary as in thermodynamics. This
allows to unify the formalisms and transpose the results obtained in
one context to another context. Indeed, we have found many {\it
analogies} between systems of a very different nature: stellar
systems, vortices, bacteria, Bose-Einstein condensation...
\cite{gfp,houches,iso,bose}.




\end{document}